\documentclass[12pt]{iopart}

\usepackage{graphicx}

\begin{document}

\title[Strong energy-momentum dispersion in the lightly doped SrTiO$_3$]{Strong energy-momentum dispersion of phonon-dressed carriers in the lightly doped band insulator SrTiO$_3$}

\author{W. Meevasana$^{1, 2, 3, 4}$, X. J. Zhou$^{5}$, B. Moritz$^{2}$, C-C. Chen$^{1, 2}$, R.H. He$^{1, 2}$, S.-I. Fujimori$^{6}$, D.H. Lu$^{2}$, S.-K. Mo$^{1, 7}$, R.G. Moore$^{2}$, F. Baumberger$^{3}$, T.P. Devereaux$^{2}$, D. van der Marel$^{8}$, N. Nagaosa$^{9,10}$, J. Zaanen$^{11}$, Z.-X. Shen$^{1,2}$}

\address {$^1$Departments of Physics and Applied Physics, Stanford
University, CA 94305, USA}

\address{$^2$Stanford Institute for Materials and Energy Sciences,
SLAC National Accelerator Laboratory, 2575 Sand Hill Road, Menlo
Park, CA 94025, USA}

\address{$^3$School of Physics and Astronomy, University of
St Andrews, North Haugh, St. Andrews, Fife KY16 9SS, UK}

\address{$^4$Synchrotron Light Research Institute, Nakhon Ratchasima
30000, Thailand}

\address{$^5$Institute of Physics, Chinese Academy of
Sciences, Beijing 100190, China}

\address{$^6$Synchrotron Radiation Research Unit, Japan
Atomic Energy Agency, Mikazuki, Hyogo 679-5148, Japan}

\address{$^7$Advanced Light Source, Lawrence Berkeley
National Lab, Berkeley, CA 94720, USA}

\address{$^8$D\'epartement de Physique de la Mati\`ere
Condens\'ee, Universit\'e de Gen\`eve, quai Ernest-Ansermet 24,
CH1211 , Gen\`eve 4, Switzerland}

\address{$^9$Department of Applied Physics, The University of
Tokyo, Bunkyo-ku, Tokyo 113–8656, Japan}

\address{$^{10}$Cross-Correlated Materials Research Group
(CMRG), ASI, RIKEN, Wako 351-0198, Japan}

\address{$^{11}$The Instituut-Lorentz for Therorectical
Physics, Leiden University, Leiden, The Netherlands}

\begin{abstract}
Much progress has been made recently in the study of the effects
of electron-phonon (el-ph) coupling in doped insulators using
angle resolved photoemission (ARPES), yielding evidence for the
dominant role of el-ph interactions in underdoped cuprates. As
these studies have been limited to doped Mott insulators, the
important question arises how this compares with doped band
insulators where similar el-ph couplings should be at work. The
archetypical case is the perovskite SrTiO$_3$ (STO), well known
for its giant dielectric constant of 10000 at low temperature,
exceeding that of La$_2$CuO$_4$ by a factor of 500. Based on this
fact, it has been suggested that doped STO should be the
archetypical bipolaron superconductor. Here we report an ARPES
study from high-quality surfaces of lightly doped SrTiO$_3$.
Comparing to lightly doped Mott insulators, we find the signatures
of only moderate electron-phonon coupling: a dispersion anomaly
associated with the low frequency optical phonon with a
$\lambda'\sim0.3$ and an overall bandwidth renormalization
suggesting an overall $\lambda'\sim0.7$ coming from the higher
frequency phonons. Further, we find no clear signatures of the
large pseudogap or small polaron phenomena. These findings
demonstrate that a large dielectric constant itself is not a good
indicator of el-ph coupling and highlight the unusually strong
effects of the el-ph coupling in doped Mott insulators.
\end{abstract}

\submitto{\NJP}
\maketitle

\section{Introduction}
The notion that carriers doped into insulators get dressed by
lattice deformations has been around for a long
time\cite{review:Devreese, review:Alexandrov}. A recent
development is that this polaron formation can be studied
experimentally using ARPES yielding more direct information on the
physics than classical transport and optical spectroscopic
methods. Especially when the carrier density is small but finite,
where a controlled theoretical framework is lacking, ARPES has
been quite revealing. The case has been made that lightly doped
cuprates fall victim to small polaron formation (strong
interacting case) that is vulnerable to self trapping by
impurities \cite{Polaron:Kyle,LSCO:Xingjiang}: in undoped cuprates
the spectral functions reveal Frank-Condon type broad humps caused
by the coupling to multiple phonons, and only when doping is
increased, a well-defined quasi-particle (QP) peak starts to
emerge \cite{Polaron:Kyle, LSCO:Xingjiang}. Another recent ARPES
revelation is found in the context of highly doped manganites in
the colossal magneto resistance regime. At high temperatures ARPES
reveals the Frank-Condon humps signaling small polarons, while
upon lowering temperature small pole-strength quasiparticle peaks
appear in addition, indicating that a coherent Fermi-liquid is
formed from the microscopic polarons\cite{LSMO:Norman}.

Both manganites and cuprates are doped Mott-insulators and no
modern ARPES information is available on polaron physics in the
simpler doped band insulators. We therefore decided to focus on
the classic SrTiO$_3$ doped band insulator. SrTiO$_3$ is known to
have an exceptionally high static dielectric constant on the order
of $10^4$ at low temperature \cite{incipient:Muller}.
Superconductivity can be induced by electron doping with either O,
Nb, or La\cite{Tc:Koonce,Superconducting:Suzuki} over a narrow
range of low carrier concentrations between $\sim 10^{19}$ to
$\sim 10^{20}$ cm$^{-3}$. The optimal T$_{c}$ is typically 0.2-0.3
K but can reach up to 1.2 K\cite{HighTc:Bednorz} which is
surprisingly high for such low carrier concentrations. It has been
speculated that this is due to the formation of
bipolarons\cite{bipolaron:Devreese}. However, whether large or
small polarons actually exist in SrTiO$_3$ depends on the relevant
length scale for the electron-phonon couplings.

The case was made in a recent optical study by van Mechelen
\emph{et al.} that the electron-phonon coupling is actually not
very strong\cite{optical:vandermarel}. ARPES is however more
direct in revealing the strength of  the coupling to specific
phonons. With this technique we arrive at the conclusion that
small polarons are not formed in STO and that the electron-phonon
coupling acts in a perturbative way.

\section{Methods and Materials}

The samples investigated here are
La$_{x}$Sr$_{1-x}$TiO$_{3+\delta}$ (Crystal Base Co., Japan) at
nominal dopings of $x=0.01$ (T$_c$ $\sim$ 0.2 K) and $x=0.05$ (non
superconducting) \cite{Superconducting:Suzuki} while the actual
doping levels at the surface are slightly different due to oxygen
vacancies. We obtain high-quality surfaces by cleaving along
guiding lines at T = 10 K and measure at the same temperature.
This new technique results in significantly flatter surfaces than
fracturing or scraping of SrTiO$_3$. This was found to
substantially improve the quality of ARPES data and enable us to
see a clear quasi-particle band dispersion and dispersion anomaly
which have not been seen in previous
measurements\cite{MottInsulator:Review,Screening:Shin,LaSTO:Aiura}

ARPES data were collected on a Scienta-4000 analyzer at the
Stanford Synchrotron Radiation Laboratory (SSRL), Beamline 5-4,
and the Advanced Light Source (ALS), Beamline 10.0.1, with photon
energies between 18-90 eV and a base pressure of $< 4 \times
10^{-11}$ torr. Samples were cleaved \emph{in situ} along the
(001) plane at the measurement temperature, T = 10 K. A sharp (1
$\times$ 1) low-energy-electron-diffraction pattern indicates a
well-ordered surface devoid of any reconstructions. The energy
resolution was set to 9-11 meV and 15-20 meV for 18-35 eV and
35-90 eV photon energies respectively and the angular resolution
was ~0.35$^{\circ}$. Additionally, a LSCO sample with x=0.01 was
measured at ALS with photon energy = 55eV and T = 20K.

\section{Results}

\begin{figure} [t]
\linespread{1}
\par
\begin{center}
\hspace*{-0.5cm}
\includegraphics [width=4.5in, clip]{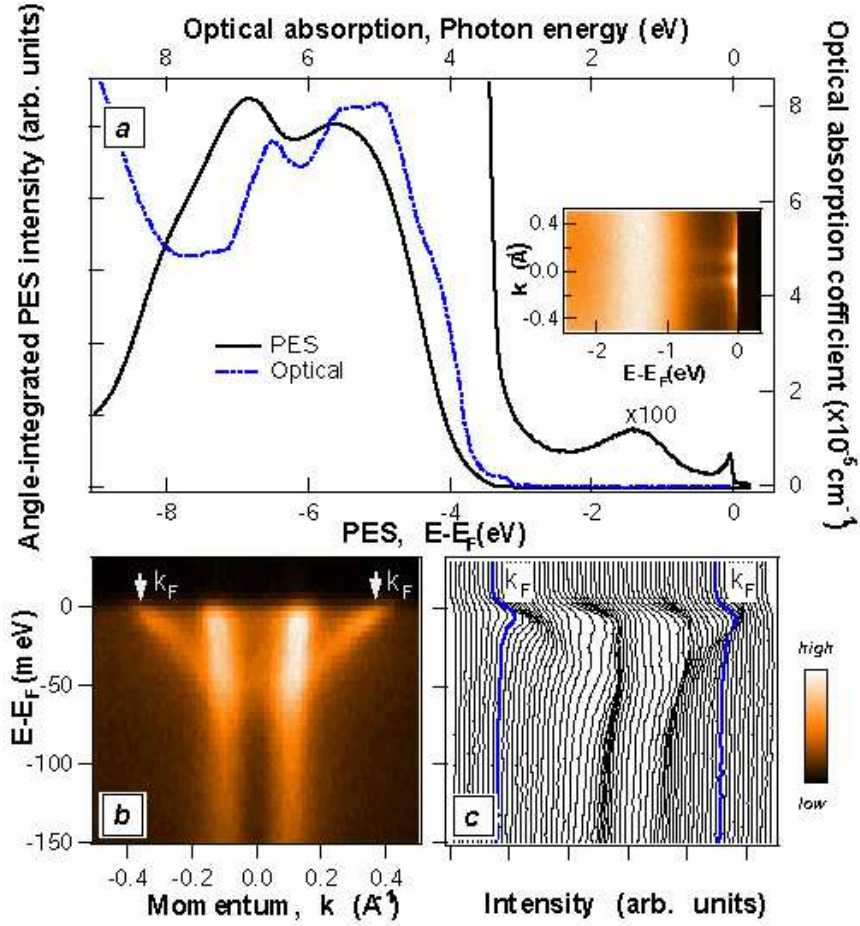}
\end{center}
\caption{\label{fig1} ARPES data of $x=0.01$ sample at $T = 10$K.
(a) Angle-integrated photoemission spectrum up to 9 eV in binding
energy, together with optical absorption data of an undoped sample
from Ref. \cite{Gap:Cardona}. The inset shows angle-resolved data
of the in-gap state around 1.3 eV. (b) Quasi-particle band
dispersion in the (010) plane near E$_F$ (see cut b in Fig. 2(f) )
with corresponding energy distribution curves in (c).}
\end{figure}

\begin{figure*}
\linespread{1}
\par
\begin{center}
\hspace*{-0.5cm}
\includegraphics [width=4.5in, clip]{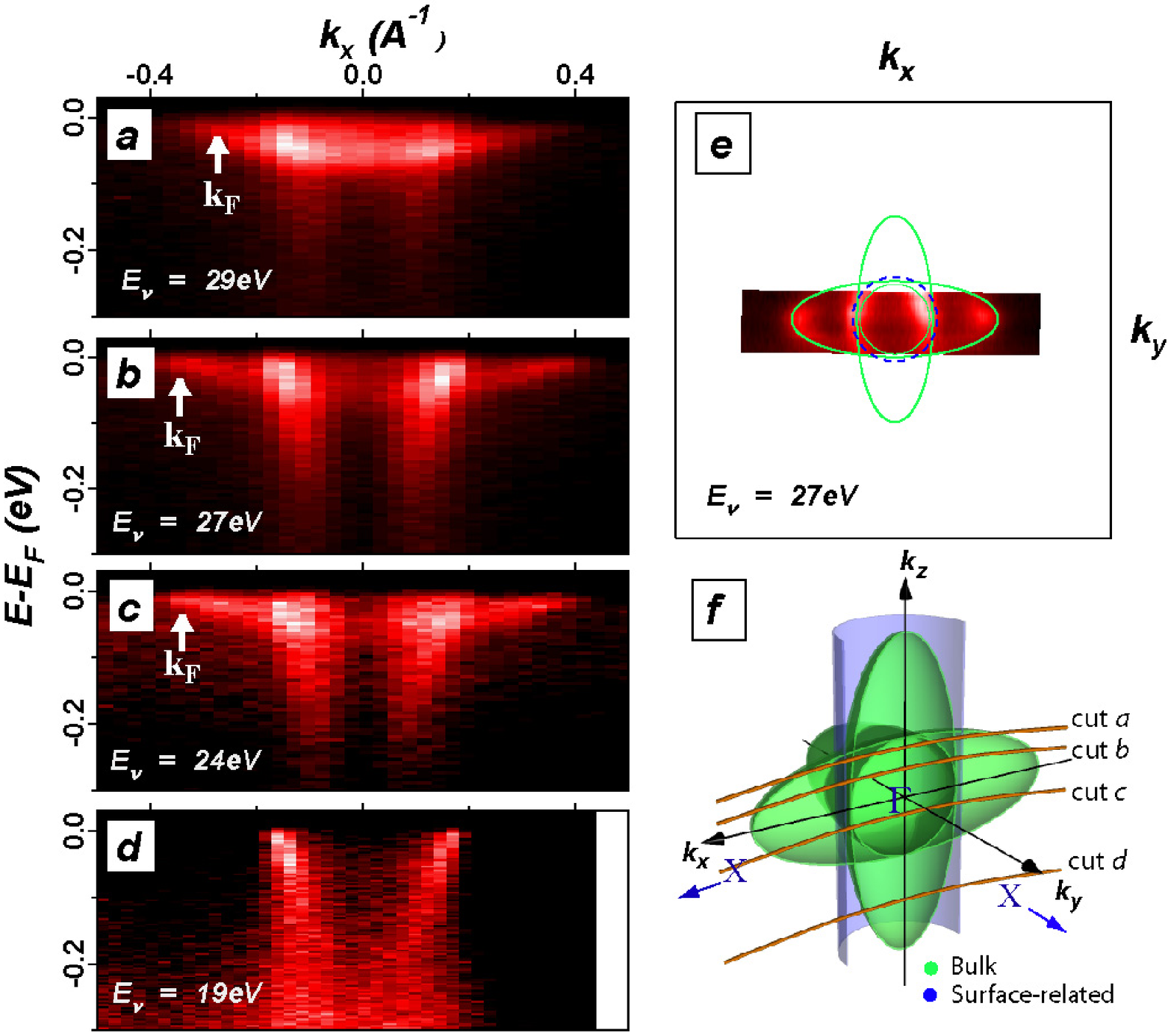}
\end{center}
\caption{\label{fig2:kz} Fermi surface topography of SrTiO$_3$.
(a) - (d) Band dispersion in the (010) plane for photon energies
of 19, 24, 27 and 29 eV, respectively. Doped SrTiO$_3$ has a cubic
unit cell and a three-dimensional Fermi surface, consisting of
three ellipsoid-like surfaces along each axis (see more detail in
Ref. \cite{LaSTO:Aiura}); when including the spin orbit coupling
term in calculation \cite{optical:vandermarel}, a shift in the
Fermi surface occurs as shown in Fig. 3(e). (e) Fermi surface map
near the Brillouin zone mid-plane ($h\nu =$27eV). The solid green
lines are guides to the eye. Estimated $k_z$ positions for (a)-(d)
are indicated by the orange lines in the schematic Fermi surface
(f), where half of the whole Fermi surface is plotted; green
(blue) indicates bulk (surface) band. Note that in $x=0.01$
samples a second bulk band is expected with a Fermi crossing near
the surface band. However, this band appears to be overshadowed by
the more intense surface related band and further suppressed by
the matrix element near $\Gamma$ point.}
\end{figure*}

In Fig. 1, we present ARPES data taken at a photon energy of 27
eV. The dominant features in the angle integrated spectrum are the
valence band between 3.3 eV and 9 eV, an in-gap state near 1.3 eV
and the QP peak at the Fermi level. The energy gap between the
onset of the oxygen valence bands to the QP band bottom is around
3.3 eV consistent with optical measurements\cite{Gap:Cardona}
while local-density-approximation (LDA) band structure
calculations predict a gap of $\sim2$ eV\cite{LaSTO:Aiura,
LDA:Cohen}. The presence of a non-dispersive and broad in-gap
state around 1.3 eV has been discussed in the literature (Ref.
\cite{Screening:Shin} and refs. therein) as caused by a local
screening effect, chemical disorder or donor levels.

Having established the basic spectral features, we now focus on
the Fermi surface topography of SrTiO$_3$. Fig. 2(a)-2(d) show
ARPES data taken at various photon energies (changing k$_z$)
together with a Fermi surface map at 27 eV, projected on the
$k_x$-$k_y$ plane (Fig. 2(e)). The flatter band with a $\sim$60
meV band bottom (i.e. in Fig. 2(a)-2(c)) corresponds to a bulk
state since the $k_F$ crossing changes with different photon
energy (changing $k_z$), in agreement with LDA calculations by
I.I. Mazin where the computational details are the same as in Ref.
\cite{optical:vandermarel}. The steeper band with a bottom
$\sim$200 meV (e.g. in Fig. 2(d)) can be attributed to the surface
of cleaved SrTiO$_3$ because the data do not show noticeable
dispersion along $k_z$ and they are absent in LDA calculations
(indicated by blue line and surface in Fig. 2(e) and 2(f)). Since
the surface band crosses the bulk band, it cannot be an eigenstate
of the system but possibly a surface resonance state; this surface
state is investigated further in smaller doping, x=0.001 samples
\cite{STO_SS:non}. We also note that in $x=0.01$ samples a second
bulk band is expected with a Fermi crossing near the surface band.
However, this band appears to be overshadowed by the more intense
surface related band and further suppressed by the matrix element
near $\Gamma$ point. In the following, we will use the schematic
contours of the $k_F$ positions, indicated by green surfaces in
Fig. 2(f), to describe the bulk Fermi surface.

\begin{figure*}
\linespread{1}
\begin{center}
\hspace*{-0.5cm}
\includegraphics [width=6in, clip]{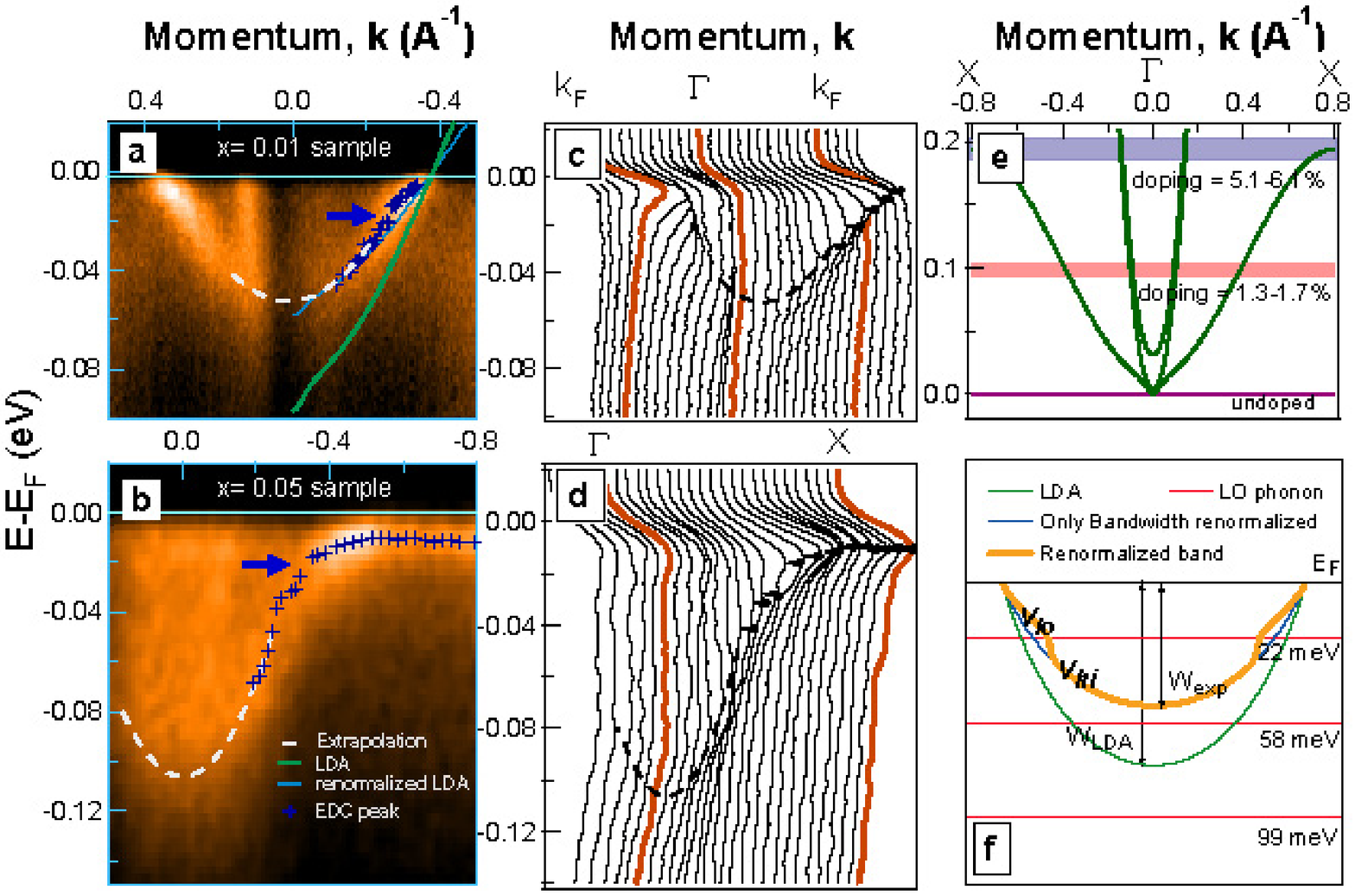}
\end{center}
\caption{\label{fig3:qp} ((a), (b) Quasi-particle band dispersions
of x=0.01 and 0.05 samples. Cross symbols indicate peaks from
energy distribution curves (EDC) and dashed lines are extrapolated
to get the band bottom. The most prominent kink energy is
indicated by arrows around $\sim20meV$ for both doped samples.
Possibly, there is a second kink in the x=0.05 sample between
40-60 meV but the complication from the side band makes it less
clear. Note that k$_z$ varies slightly along the $k$-axis of (a)
and (b) (see Fig. 2). However, this change has only a minute
influence on the measured group velocities and will be neglected
for the discussion of the mass renormalization. (c) and (d) show
EDCs of ARPES data shown in (a) and (b) respectively where the EDC
peak positions are marked by triangle symbols and the dash lines
are extrapolations. (e) LDA band dispersion of undoped SrTiO$_3$
along $\Gamma$X\cite{optical:vandermarel}. Fermi levels positions
for dopings = 1.3-1.7\% and 5.1-6.1\% are indicated by shaded
areas. (f) Schematic plot of renormalized band dispersion in the
forms of ``kink" and reduced bandwidth caused by phonons whose
mode energies are lower and higher than the electron bandwidth,
respectively.}
\end{figure*}

Fig. 3(a) and 3(b) show the occupied bands, along the $\Gamma$X
direction in the vicinity of the $\Gamma$ point of the $x=0.01$
and 0.05 samples. By aligning the $k_F$'s of the ARPES data with
those of the LDA dispersions\cite{optical:vandermarel}, we
estimate the dopings of the $x=0.01$ and $x=0.05$ samples to be
slightly higher than the nominal dopings (1.5$\pm$0.2\% and
5.6$\pm$0.5 \%, respectively: Fig. 3(e)), likely due to a small
oxygen deficiency at the surface.

Having isolated the occupied part of the conduction bands (Fig.
3(a)-3(d)), let us now turn to the interpretation of the data. In
the data one can discern a weak kink in the dispersion at
approximately 20 meV binding energy (blue arrows, Fig. 3(a) and
3(b)). This is more clear in the $x=0.01$ sample, since in the
x=0.05 sample it resides in a region where the dispersion has a
strong curvature (Fig. 3(e)). Such a weak kink structure in the
dispersion indicates a perturbative coupling with a bosonic mode
at this energy. This interpretation is supported by the
observation that the intensity rapidly increases below the kink
energy; above the kink energy, an extra decay channel opens up
that will smear the QP peak. To quantify the coupling to this
boson, we extracted the band velocities for the $x=0.01$ case at
binding energies below ($v_{lo}$) and above ($v_{hi}$) the kink
energy (see Fig. 3(f)) to be $\sim$0.16 eV${\AA}$ and 0.21
eV${\AA}$, respectively. The mass renormalization is therefore
$v_{hi}/v_{lo} =m^*/m\sim$1.3, indicating a coupling to this
particular boson $\lambda' \equiv m^*/m-1$ = 0.3.

Given that the signals are rather smeared at higher energies we
cannot exclude the presence of other kinks associated with higher
energy modes. However, the data permit us to track the overall
width of the occupied parts of the conduction bands. For the
$x=0.01$ sample we find the band bottom at $\sim$58 meV whereas
the LDA calculation indicates it to be at $\sim$97 meV
\cite{optical:vandermarel} (Fig. 3(e)). It follows that the
overall width of the occupied band is renormalized by a factor of
$\sim$1.7 (W$_{LDA}$/W$_{exp}$). The total mass renormalization is
the product of the bandwidth and kink renormalization factors and
we find this to be $1.7\times1.3 \simeq2.2$ in the $x=0.01$
sample, close to the estimate 2-3 deduced from the optical
measurements\cite{optical:vandermarel}.

To compare with STO data, Fig. 4 shows ARPES data of
La$_{1-x}$Sr$_{x}$CuO$_{4}$ with x=0.01, as a lightly doped Mott
insulator. The momentum is along the (0,0) to ($\pi, \pi$)
direction. We note that all the data in Fig. 4 are already
subtracted by the non-dispersive background of oxygen valance band
for clearer comparison. In contrast to STO, the spectrum (see red
line in Fig. 4(b)) shows a small quasi-particle peak with large
Frank-Condon type broad hump around 400 meV - a signature of small
polaron formation. LSCO data also shows a clear kink in the
dispersion, indicating a strong electron-phonon coupling at around
70 meV (see arrow in Fig. 4(a)); to quantify this coupling, we
extracted the band velocities at binding energies below ($v_{lo}$)
and above ($v_{hi}$) the kink energy to be $\sim$1.66 eV${\AA}$
and 6.14 eV${\AA}$, respectively. Therefore, the mass
renormalization factor from this kink feature is
$v_{hi}/v_{lo}=m^*/m \sim$3.7. A use of $\lambda' \equiv m^*/m-1$
would give $\lambda'$ of 2.7, giving a clear contrast to the
extracted value of $\sim$0.3 from the kink feature of the x=0.01
STO sample.

\begin{figure}[h]
\linespread{1}
\begin{center}
\hspace*{-0.5cm}
\includegraphics [width=4.5in]{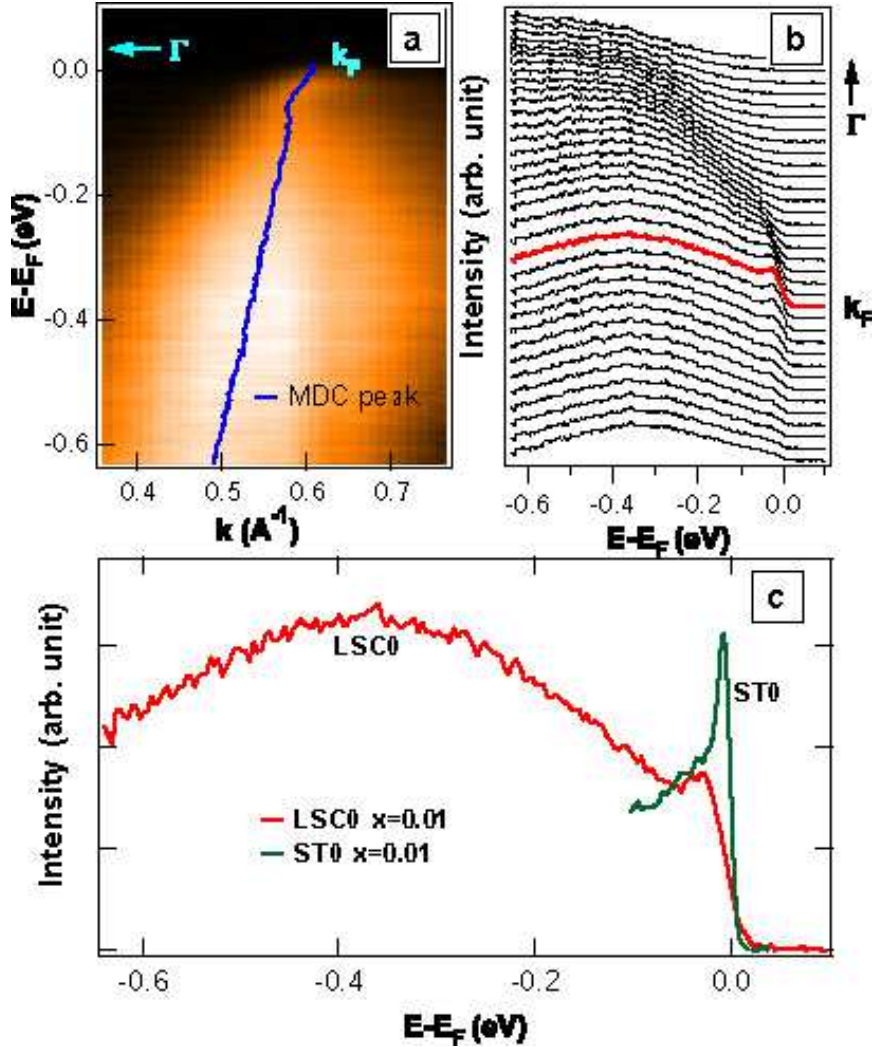}
\end{center}
\caption{ARPES data of La$_{1-x}$Sr$_{x}$CuO$_{4}$ with x=0.01.
(a) shows the raw ARPES data; blue line indicates the peak
position in momentum distribution curves (MDC) and orange circles
indicate the peak positions in energy distribution curves (EDC)
where the big arrow indicates the kink in dispersion at binding
energy $\sim$70meV. (b) shows the corresponding energy
distribution curves. (c) Comparison of ARPES spectra with
background subtracted at k$_F$ of 1) x=0.01 STO sample along
$\Gamma-X$ direction and 2) 1\% doping La$_{2-x}$Sr$_{x}$CuO$_{4}$
along (0,0) to ($\pi, \pi$) direction.}\label{Fig. 4}
\end{figure}

\section{Discussion and Conclusion}
How to interpret these findings from STO data? The 20 meV kink is
certainly related to a phonon. A-priori one can be less certain
about the cause of overall bandwidth renormalization because an
electronic origin cannot be excluded. However, although LDA is
well known to underestimate band gaps in band-insulators, it does
not usually underestimate the bandwidths and our extracted
renormalization factor may be regarded as an upper value. At the
same time, the phonon dispersions of STO have been measured by
infrared and Raman spectroscopy\cite{Gap:Cardona} and neutron
scattering\cite{neutron:Cowley,Shirane1, Shirane2,
phonon:Choudhury} in great detail; the phonon modes are in the
range of 0-100 meV where much of the phonon spectrum extends to
energies that are larger than the Fermi energy of at least the
$x=0.01$ system. Under such an anti-adiabatic condition, one
expects the electron-phonon coupling to give rise to an overall
bandwidth renormalization that can be estimated from the
mass-renormalization formulae for the isolated polaron
\cite{review:Devreese}. Since the focus here is on the
surprisingly moderate el-ph coupling, we attribute all the
renormalization to el-ph interaction which sets the upper bound
for the value of $\lambda \sim 1$. An overall coupling $\lambda
\sim 1$ can mean both that small, self-trapped polarons are formed
but also that the system stays itinerant. What decides the
nomenclature is the length scale of the relevant electron-phonon
couplings.

When the el-ph coupling is short ranged, small polarons are
expected. One can take the cuprates as an example where an
effective $\lambda \simeq 1$ corrected for electronic band
narrowing effects \cite{polaron:Gunnarsson} that enhance the
impact of el-ph interaction is believed to be responsible for the
multi-phonon Franck-Condon peak indicated in Fig. 4(c). Here we
should note that m* is no longer linear with $\lambda$ and
increases rapidly near the small-to-large polaron crossover around
$\lambda \simeq 1$. For $\lambda \simeq 1$ in cuprates (e.g. in
the case of LSCO shown in Fig. 4), the actual face value of mass
renormalization could be as large as 3.7; hence, $\lambda' $
defined by $m^*/m-1$ would be 2.7.

The most striking aspect of the STO data is that such effects due
to small-polaron formation are entirely absent in STO, where
instead the electrons remain strongly coherent as manifested by
the strong energy-momentum dispersion and the rather sharp QP
peaks with large pole strengths even for 1\% doped sample (Fig.
4(c)). This can be reconciled with the relatively large $\lambda$
assuming that the dominating electron-phonon couplings are of the
long ranged, polar kind \cite{Toyozawa}. This claim can in fact be
further substantiated by the finding that our data are in
semi-quantitative agreement with `naive' continuum limit
estimations of the polar el-ph interactions\cite{polaron:EaglesDM,
bipolaron:Devreese}. In this way only the long ranged
electrostatic interactions are taken into account with the
longitudinal optical (LO) phonons, omitting completely short range
interactions involving the transversal optical (TO) phonons that
are in reality always present.

Starting from this perspective, let us first discuss why the large
bulk dielectric constant may not be a good indicator for formation
of small polaron. The dielectric constants at zero frequency
($\varepsilon_0$) and at frequencies large compared to the phonon
energy ($\varepsilon_\infty$) are related to the frequencies of LO
and TO phonons as $\varepsilon_0/\varepsilon_{\infty} = \Pi_a
(\omega_{a LO}/\omega_{aTO})^2$ where $a$ specifies the phonon
branch. A large $\varepsilon_0$ signals a softening of the TO
phonon that eventually can condense in a ferroelectric state.
However, the Fr$\ddot{o}$hlich polar el-ph interactions involve
the LO phonons where the large $\varepsilon_0$ will also help the
coupling strength but with smaller effect. The coupling strength
depends on the dielectric term $\kappa^{-1} =
\varepsilon_{\infty}^{-1} - \varepsilon_0^{-1}$ which increases
slightly upon increase of $\varepsilon_0$ ($\kappa^{-1}\sim0.17$
in La$_2$CuO$_4$ and $\sim 0.19$ in STO) \cite{kappa:Alexandrov}.
Another issue is that the short-range coupling to this TO phonon
could be enhanced due to the softening of the frequency
$\omega_{TO}$. However, since the softening occurs only in a
narrow region in momentum space characterized by the scale $a/\xi
\cong 0.1$ ($\xi$: correlation length)~\cite{Shirane1, Shirane2},
the increase in the coupling constant $\Delta \lambda$ of the
order of $\Delta \lambda \cong \lambda
(\omega^0_{TO}/\omega_{TO})^2(a/\xi)^3$ is small.

\begin{table}[h]
\caption{\label{table1:node} Comparison of features between
SrTiO$_3$ and cuprates - perovskite band and Mott insulators.}
\begin{indented}
\item[]\begin{tabular}{ccc}
Feature& Mott Insulator&Band
Insulator\\
&(La$_2$CuO$_4$)&(SrTiO$_3$)\\ \hline
Mass renormalization factor from\\
kink feature at small doping x=0.01&$\sim$3.7&$\sim$1.3\\\\
Small polaronic effect\\at small doping&Yes&No\\\\
Large pseudogap behavior\\at small doping&Yes&No\\\\
Small Fermi Surface \\pocket at small doping&Maybe(YBCO)\cite{Fermipocket:Proust}&Yes\\\\
Dielectric constant \\(undoped)&$\sim 20$\cite{LCO:Tamasaku}&$\sim 10^2-10^4$\\\\
\hline
\end{tabular}
\end{indented}
\end{table}

Under these assumptions one is, according to the calculations of
Devreese et al.\cite{polaron:EaglesDM, bipolaron:Devreese}, left
with three LO phonons at (for $q=0$) 22, 58, and 99 meV with
coupling constants $\alpha_i$ of 0.018, 0.945 and 3.090,
respectively. Using that, for weak coupling, $\lambda_i =
\alpha_i/6$ \cite{review:Devreese, Feynman} this translates into
$\lambda_i$'s of 0.003, 0.16 and 0.6 respectively. These modes are
indicated together with the electronic dispersions in the
schematic Fig. 3(f). The low energy kink in the electron
dispersions matches very well with the 22 meV mode associated with
Sr-O bond stretching \cite{phonon:Choudhury}. The other two phonon
modes are at higher energy than the band bottom and hence they
should cause an overall band width reduction. For an accurate
treatment, one may consider the anti-adiabatic limit; however from
the available calculation, the coupling constants $\alpha$ of
these 58-meV and 99-meV modes will give a bandwidth-renormalized
factor of 1.76, which is already very close to our extracted value
of $\sim$1.7. The polar el-ph calculation strongly underestimates
the $\lambda' \simeq 0.3$ coupling to the 22 meV phonon. The main
coupling from this Sr-O bond stretching modes comes from large
momenta near the zone boundary as in the cuprates and is expected
from general grounds due to the displacement eigenvectors. Thus,
it is a local deformation. As we discussed in the previous
paragraph this could well be significantly enhanced by the
proximity to the ferroelectric transition, a reason why its main
impact is only on the low energy phonon. However, given that the
coupling is still moderate and this phonon is of the
adiabatic/Eliashberg kind, it does not interfere with the
consistency of our argument.

While small polarons are absent in STO, the cuprates at similar
doping show a sharp contrast in displaying strong el-ph coupling
with mass renormalization as large as 3.7 (see Table 1 for the
comparison between LSCO and SrTiO$_3$). For the following reasons,
the carriers doped into a Mott insulator can be subject to a
stronger short-range el-ph interaction. One is that the additional
polaronic effect due to the magnetic degrees of freedom enhances
the effective mass, and hence collaborate to form the composite
small polaron with magnon and phonon clouds\cite{Mishchenko}.
Another reason is that fluctuations with large momentum (e.g.
k=$(\pi/2,\pi/2)$ of the antiferromagnetic state in the cuprates)
are involved in dressing the doped carriers. Starting from the
Fr$\ddot{o}$hlich interaction, the exchange of this large momentum
can lead to the short-range el-ph interaction. In the case of a
band insulator, the large momentum in this same order of magnitude
is not immediately available. Therefore, small polaron formation
is more likely to occur in a Mott insulator than in a band
insulator. We should note that there could be an additional
advantage in anisotropic layer compounds that electron-phonon
coupling along perpendicular axis is little screened and hence
remains strong\cite{Screening:Alexandrov, Screening:non}. However,
it is also known in cuprates that the small polaron effect
disappears largely upon doping away from the antiferromagnetism
(e.g. Na$_x$Ca$_{2-x}$CuO$_2$Cl$_2$ with x=0.12
\cite{Polaron:Kyle}) where the conductivity is still very
anisotropic. Therefore, it is likely that there are also other
physics (e.g. as discussed above) in helping the formation of
small polaron.

In conclusion, we have shown the quite unexpected results that
there is little evidence for small polaron formation in lightly
doped SrTiO$_3$ indicating that the large dielectric behavior can
occur independently of strong el-ph interactions. In turn, this
indicates that in doped Mott insulators like the cuprates, the
dressing of electrons by spin excitations
\cite{polaron:Gunnarsson, Mishchenko} and strong correlations
conspire to give a short-range el-ph interaction able to trap
doped carriers and more readily form polarons.

\section{Acknowledgments}
We gratefully thank II Mazin for providing the unpublished LDA
calculations, A Mishchenko for helpful discussion and H Takagi and
J Matsuno for crystal information. This work was supported by the
Department of Energy, Office of Basic Energy Sciences under
contract DE-AC02-76SF00515. WM acknowledges The Thailand Research
Fund for financial support. CCC was supported in part by National
Science Council, Taiwan, under grant no.
NSC-095-SAF-I-564-013-TMS.

\end{document}